\documentclass[useAMS,usenatbib]{mn2e}
\pdfoutput=1
\usepackage[version=3]{mhchem}

\usepackage{natbib}
\usepackage{graphicx}
\usepackage{stmaryrd} 
\usepackage[makeroom]{cancel}
\usepackage{lastpage}

\title[Low-temperature chemistry of \ce{H2O} (\ce{D2O}) and OD (OH)]
  {Low-temperature chemistry between water and hydroxyl radicals: H/D isotopic effects}
\author[T.~Lamberts et al.]
  {T.~Lamberts$^{1,2}$\thanks{corresponding author, current address: Computational Chemistry Group, Inst. of Theoretical Chemistry, University of Stuttgart, Pfaffenwaldring 55, D-70569, Stuttgart, Germany }, G.~Fedoseev$^1$, F.~Puletti$^3$, S.~Ioppolo$^{2}$, H.~M.~Cuppen$^2$, and H.~Linnartz$^1$\\  
  $^1$~Raymond and Beverly Sackler Laboratory for Astrophysics, Leiden Observatory, University of Leiden, \\\hspace{10pt}P.O. Box 9513, NL 2300 RA Leiden, The Netherlands.\\
  $^2$~Theoretical Chemistry, Institute for Molecules and Materials, Radboud University Nijmegen, \\\hspace{10pt} Heyendaalseweg 135, 6525 AJ Nijmegen, The Netherlands \\
  $^3$~Department of Chemistry, University College London,\\\hspace{10pt} 20 Gordon Street, London WC1H 0AJ, UK}
\date{Released 2014 Xxxxx XX}

\pagerange{\pageref{firstpage}--\pageref{lastpage}} \pubyear{2014}

\def\LaTeX{L\kern-.36em\raise.3ex\hbox{a}\kern-.15em 
    T\kern-.1667em\lower.7ex\hbox{E}\kern-.125emX}

\begin{document}

\label{firstpage}

\maketitle

\begin{abstract}

Sets of systematic laboratory experiments are presented - combining UHV cryogenic and plasma-line deposition techniques - that allow to compare H/D isotopic effects in the reaction of \ce{H2O} (\ce{D2O}) ice with the hydroxyl radical OD (OH). The latter is known to play a key role as intermediate species in the solid state formation of water on icy grains in space. The main finding of our work is that the reaction \ce{H2O + OD -> OH + HDO} occurs and that this may affect the \ce{HDO}/\ce{H2O} abundances in space. The opposite reaction \ce{D2O + OH -> OD + HDO} is much less effective, and also given the lower \ce{D2O} abundances in space not expected to be of astronomical relevance. The experimental results are extended to the other four possible reactions between hydroxyl and water isotopes and are subsequently used as input for Kinetic Monte Carlo simulations. This way we interpret our findings in an astronomical context, qualitatively testing the influence of the reaction rates.

\end{abstract}


\section{Introduction}
Surface reactions on grains have been proposed as an effective way to form water at the low temperatures typical for the interstellar medium  \citep{Hulst:1949, Tielens:1982}. Over the last ten years many  of the possible reactions have been tested in various laboratories and reaction routes, rates, and branching ratios have been determined \citep{Hiraoka:1998,  Miyauchi:2008, Ioppolo:2008, Ioppolo:2010, Cuppen:2010B, Romanzin:2011}. The general conclusions obtained in these studies are in line with each other \citep{Dulieu:2011IAU} and the astronomical relevance has been summarized by \citet{Dishoeck:2013}.

A detailed study of the solid state formation of water also holds the potential to characterize deuteration effects. With the goal to understand how water was delivered to Earth, there has been much interest in linking the \ce{HDO}/\ce{H2O} ratio in cometary, interstellar, and laboratory ices as well as astronomical (gas-phase) observations to ratios as found in our oceans \citep{Rodgers:2002, Hartogh:2011}. Therefore, both the origin of the \ce{HDO}/\ce{H2O} ratio in the ices and their subsequent chemical and thermal processing are currently widely studied \citep{Caselli:2012}. It is in fact during the water formation on dust surfaces when the deuterium fractionation commences. The preferential incorporation of D over H in molecules can lead to D/H ratios in molecules that are much larger than the primordial ratio of $\sim 1.5\times10^{-5}$ \citep{Piskunov:1997, Oliveira:2003}. To understand the origin of this fractionation, it is neccessary to consider both the gas-phase and solid-state processes that are at play. The driving forces behing these processes are the lower zero-point energy of an X-D bond with respect to an X-H bond and the difference in tunneling behavior, both as a result of the larger mass of deuterium \citep{Roberts:2003, Tielens:1983, Caselli:2012, Lipshtat:2004}. One of the possible enhancement routes for hydrocarbon bonds is the simple replacement of a hydrogen by a deuterium \emph{via} a deuterium mediated abstraction mechanism: \ce{- CH + D -> - C + HD} followed by \ce{- C + D -> - CD} \citep{Nagaoka:2005, Nagaoka:2007}.  An example is substitution of H for D in solid methanol, which has been found to be efficient, whereas the reverse reaction (substituting D for H) is not. Note that \citet{Nagaoka:2005} have also attempted to substitute hydrogen in water by the same process, but did not find any deuteration upon D exposure. The non-occurence of reaction \ce{H2O + D} has been confirmed in our laboratory (unpublished data).

Another abstraction process, so far studied in less detail, involves OH and OD radicals. Hydroxyl radicals play an important role as reactive intermediates in water formation (see e.g. \citealp{Cuppen:2010B}). Furthermore, recent microscopic models have shown that the radical concentration in the ice can be very high if photon penetration is included \citep{Chang:2014}. \citet{Garrod:2013} indicated the importance of abstraction reactions by the hydroxyl radical in the framework of complex hydrocarbon molecules. Such an OH induced abstraction mechanism is particularly important in water-rich ices, because both OH and OD radicals are produced on the surface by radical chemistry and by photodissociation of water isotopologues \citep{Ioppolo:2008, Andersson:2008, Oberg:2009B}. 

Considering again the formation of \ce{HDO} in the ice, the water surface reaction network needs to be duplicated, involving both reactions with hydrogen and deuterium. As a result of the large number of reactions that are in competition with each other and with diffusion, it is experimentally challenging to study hydrogenation and deuteration, simultaneously. Therefore, in the past, either the deuteration pathways have been studied separately, \emph{e.g.}, \ce{O2 + D} \citep{Chaabouni:2012a}, or specific reaction routes are tackled theoretically and experimentally, \emph{e.g.}, \ce{OH\;(OD) + H2\;(HD\; \text{or}\; D2)} and \ce{H2O2\;(D2O2) + H\;(D)} \citep{Kristensen:2011, Oba:2012, Oba:2014}. Here, we add to these studies and investigate the cross links between the hydrogenation and deuteration networks by the following four hydrogen abstraction reaction \emph{via} hydroxyl radicals:
\begin{align}
\ce{H2O + OD} &\xrightarrow{k_1} \ce{OH + HDO} \;\; \tag{R1}\label{R1} \\
\ce{HDO + OD} &\xrightarrow{k_2} \ce{OH + D2O} \;\; \tag{R2}\label{R2} \\
\ce{HDO + OH} &\xrightarrow{k_3} \ce{OD + H2O} \;\; \tag{R3}\label{R3} \\
\ce{D2O + OH} &\xrightarrow{k_4} \ce{OD + HDO} \;\; \tag{R4}\label{R4} \;.
\end{align}

Hydrogen abstraction of OH from \ce{H2O} (or the fully deuterated analog) can also take place
\begin{align}
\ce{H2O + OH} &\xrightarrow{k_5} \ce{OH + H2O} \;\; \tag{R5}\label{Rlast} \\
\ce{D2O + OD} &\xrightarrow{k_6} \ce{OD + D2O} \;\; \tag{R6}\label{Rlast2}
\end{align}
and although this does not have a net effect on the abundances in the ice, it can be seen as an analog of bulk diffusion, of which models indicate it is of great importance in ice chemistry \citep{Vasyunin:2013, Lamberts:2014, Chang:2014}. {The isolated OH-\ce{H2O} complex has been detected in a variety of matrices (Ar, Ne, \ce{O2}) as has been summarized by \citet{Do:2014}. This shows that a pre-reactive complex indeed can be formed.}

Reactions~\ref{R2} and~\ref{R3} cannot be studied in the laboratory, because it is not feasible to deposit pure HDO. Room-temperature rapid proton transfer reactions scramble the protons and deuterons and yield a statistical $\sim$ 1:2:1 mixture of \ce{H2O}:HDO:\ce{D2O} if one were to start from a pure HDO liquid. Such scrambling has been found to take place efficiently even in ices at temperatures far below room temperature \citep{Lamberts:2015}. Therefore, here, only reactions~\ref{R1} and~\ref{R4} are tested experimentally, at low temperature and using reflection absorption infrared spectroscopy (RAIRS) as an in-situ diagnostic tool.

Note that from an astrochemical point-of-view, reactions~\ref{R2},~\ref{R4}, and~\ref{Rlast2} are unlikely to be relevant as a result of the low concentrations of the reactants (\ce{D2O} and OD) present in the ice. Nevertheless, we put efforts in characterizing \ref{R4}, as an understanding of the underlying mechanism helps in painting the full picture. Reactions~\ref{R1} and~\ref{R3}, on the contrary, could occur in regions with a high photon flux as this causes water and isotopologues to dissociate, thus generating additional hydroxyl radicals. 
 
In the following, we outline the experimental setup and sets of experiments performed (Section~\ref{sec:ExpMeth}), the analysis of the resulting RAIR spectra (Section~\ref{sec:RnD_OHOD}), the astrochemical implications by means of a Kinetic Monte Carlo model (Section~\ref{sec:AC_OHOD}), and we conclude with summarizing remarks (Section~\ref{sec:concl_OHOD}).

\section{Experimental Methods}\label{sec:ExpMeth}
Two sets of representative experiments and their corresponding control experiments are summarized in Table~\ref{tab:exps_OHOD}. All measurements are performed at a surface temperature of 15~K  for a duration of 90~minutes. The two experiments that are used for this study are part of a larger set, varying mixing ratios and temperatures, and found to be optimum for the goals set in this work. The findings of the other experiments are largely in line with the ones discussed here, but do not provide additional information.

\begin{table}
\caption{Summary of the performed experiments with additional calibration and control experiments and corresponding parameters, $T_\text{surf} = 15$~K and $t_\text{exp} = 90$~min in all cases. The deposition rate of the species is denoted as $f_\text{dep}$ and the angle represents the angle of the deposition line with respect to the surface. }\label{tab:exps_OHOD}
\begin{tabular}{llllcl}
\hline
	&  90$^{\circ}$  & $f_{\textrm{dep}}$ 	& 45$^{\circ}$  & Discharge	& $f_{\textrm{dep}}$ 	\\
	&  		& (cm$^2$ s$^{-1}$)		& 			&  		& (cm$^2$ s$^{-1}$)		  \\
\hline\hline
1a	&   \ce{H2^{18}O}	& $4 \times 10^{12}$		& \ce{D2O} 		& $\lightning$		& $\leq 1 \times 10^{13}$ $^{a}$	  \\
1b	&   \ce{H2^{18}O}	& $4 \times 10^{12}$		& \ce{D2O} 		& -- 			& $\;\;\;\: 1 \times 10^{13}$  	 \\
1c	&   --		& --			& \ce{D2O} 		& $\lightning$		& $\leq 1 \times 10^{13}$ $^{a}$	  \\
1d	&   \ce{H2^{18}O}	& $4 \times 10^{12}$		& -- 			& -- 			& $\;\;\;\;\:$ -- 		 \\
\hline
2a	&   \ce{D2O}	& $4 \times 10^{12}$		& \ce{H2^{18}O} 	& $\lightning$		& $\leq 1 \times 10^{13}$ $^{a}$	 \\
2b	&   \ce{D2O}	& $4 \times 10^{12}$		& \ce{H2^{18}O} 	& -- 			& $\;\;\;\: 1 \times 10^{13}$  	  \\
2c	&   --		& --			& \ce{H2^{18}O} 	& $\lightning$		& $\leq 1 \times 10^{13}$ $^{a}$	 \\
2d	&   \ce{D2O}	& $4 \times 10^{12}$		& -- 			& -- 			& $\;\;\;\;\:$ -- 		  \\
\hline
\end{tabular}
$^{a}${The upper limit is derived from the \ce{D2O} and \ce{H2^{18}O} deposition rate when the microwave source is turned off.}
\end{table}

Experiments are performed using the SURFRESIDE$^2$ setup, which was constructed to systematically investigate solid-state reactions leading to the formation of molecules of astrophysical interest at cryogenic temperatures. The setup has already been extensively described in \citet{Ioppolo:2013} and therefore only a brief description of the procedure is given here.

SURFRESIDE$^2$ consists of three UHV chambers with a room-temperature base-pressure between $10^{-9}-10^{-10}$ mbar. A rotatable gold-coated copper substrate in the center of the main chamber is cooled to the desired temperature using a He closed-cycle cryostat with an absolute temperature accuracy of $\leq$ 2~K. 

Both reactions~\ref{R1} and \ref{R4} require a water isotopologue to be co-deposited along with a hydroxyl isotopologue. The former are deposited through a metal deposition line under an angle of 90$^{\circ}$ and are prepared in a separate pre-pumped ($\leq 10^{-5}$ mbar) dosing line. After undergoing several freeze-pump-thaw cycles room-temperature vapor of \ce{H2O} or \ce{D2O} can partake in the co-deposition. Secondly, the hydroxyl radicals are generated in a Microwave Atom Source (MWAS, Oxford Scientific Ltd, \citep{Anton:2000}) using a microwave discharge (300~W at 2.45~GHz) of pure water or heavy water. This discharge is located in a separate UHV beam line with an angle of 45$^{\circ}$ with respect to the surface. This beam line can be operated independently and is separated from the main chamber by a metal shutter. We cannot quantify the relative deposition rates of all fragments -- O, \ce{O2}, \ce{H}, \ce{H2}, \ce{OH}, \ce{HO2}, and \ce{H2O} (or deuterated equivalents). However, an upper limit can be derived from the \ce{H2O} deposition rate when the microwave source is turned off: $\sim 10^{13}$ cm$^2$ s$^{-1}$. Many discharge products are thus deposited onto the surface, but only one reacts with \ce{D2O} (or \ce{H2O}). First, during the calibration stage of our setup the reaction \ce{O + H2O} is found not to take place, indicating that discharge fragments that reach the surface are no longer in an excited state \citep{Ioppolo:2013}. Both \ce{H2} and \ce{O2} are inert to reactions with other non-radical species as confirmed previously \citep{Ioppolo:2010}. The reaction of H with water isotopologues has been dicussed in the Introduction and does not take place in ices for which the water molecules are hydrogen bonded \citep{Nagaoka:2005}. The reaction of \ce{HO2} with water is endergonic and has been found to occur in the gas phase only with high barriers (several thousand Kelvin) \citep{Lloyd:1974}. This leaves only the hydroxyl radical to react with a water isotopologue. 

To confirm the presence of OH among the \ce{H2O} discharge products, we co-deposited \ce{H2O} discharge dissociation products with \ce{N2} in a $\sim$ 1:20 ratio during a separate experiment (not listed in Table~\ref{tab:exps_OHOD}). The presence of the OH radical is confirmed \emph{via} its infrared absorption at 3547 cm$^{-1}$ \citep{Cheng:1988}. Furthermore, both \ce{O + H -> OH} and \ce{O2} $\xrightarrow{\text{H}}$ \ce{HO2} $\xrightarrow{\text{H}}$ \ce{OH + OH} can lead to additional hydroxyl radicals on the surface, which we expect to thermalize quickly on a picosecond timescale \citep{Arasa:2013, Meyer:2014}. Finally, our \ce{OH} radicals or rather the produced \ce{H2O2} that forms upon OH recombination on the surface, exhibits a comparable temperature dependent behavior to the \ce{H2O2} features apparent after depositing \ce{H2O} dissociation fragments from a similar microwave-discharge plasma by \citet{Oba:2011}.

A RAIR difference spectrum with respect to the background is acquired every 5 minutes up to the final time of the experiment, 90 minutes. RAIR spectra comprise a spectral range between 4000 and 700 cm$^{-1}$ with a spectral resolution of 1 cm$^{-1}$ and are averaged over 512 scans. Our region of interest lies in the 2000-1000 cm$^{-1}$ range, \emph{i.e.}, the bending modes of \ce{H2O}, HDO and \ce{D2O}: 1660, 1490, and 1250 cm$^{-1}$, respectively. Although the bands are broad, the modes are relatively well separated and can be distinguished from each other, whereas that is not the case in the stretching regions. To further enhance the peak separation between the bands of different species we made use of \ce{H2^{18}O} instead of regular \ce{H2^{16}O}. In the end, the effect was limited as in matrix isolation experiments the bending mode of \ce{H2^{18}O} is redshifted instead of blueshifted with respect to that of \ce{H2^{16}O}. In the following, we will refer to all species without mentioning explicitly the oxygen isotope at hand. Band strengths for the bending modes are typically ill-constrained and therefore we focus here on the qualitative conclusions that can be drawn.

The fluxes mentioned in Table~\ref{tab:exps_OHOD} are calculated using the following relation 
\begin{equation}
 f_{\ce{X2O}} = \frac{c_{\ce{X2O}} \;  P_{\ce{X2O}} \langle v \rangle}{4\;k_{\rm B} \; T}
\end{equation}
where $P_{\ce{X2O}}$ is the pressure, $c_{\ce{X2O}}$ is the calibration factor for the pressure gauge for the isotopologues of water, $v$ is the thermal velocity of the vapor molecules at 300 K, $k_\text{B}$ is the Boltzmann constant, and T corresponds to the (room) temperature. The calibration factors for both water and heavy water are found to be equal \citep{Straub:1998, Itikawa:2005} and we assume here that this also holds for \ce{H2^{18}O}.  

The separate experiments within a series, 1a -- 1d and 2a -- 2d (Table~\ref{tab:exps_OHOD}), are performed sequentially, \emph{i.e.}, on top of each other to make sure that the plasma conditions and deposition rates do not vary between experiments focusing on the abstraction process and the corresponding control experiments. The order in which the measurements are performed is (c)-(a)-(b)-(d). In this particular way, first the plasma is able to stabilize before experiment (c) is started. During experiment (c) only the plasma products are deposited and spectra of the accumulated products are recorded. Subsequently, the shutter of the plasma chamber is closed, while the plasma remains switched on and a new background spectrum is recorded. It is then possible to perform experiment (a), simply by opening the shutter of both the plasma chamber and the regular deposition line. At the end of this experiment, the discharge is switched off, the shutters of both beamlines are closed and another background spectrum is recorded. By opening the shutters, experiment (b) is performed, \emph{i.e.}, no plasma fragments are deposited, but rather the parent molecule itself. Finally, both shutters are closed and a background spectrum is recorded, followed by recording the deposition of the non-dissociated species, that is, experiment (d), for which only the shutter of the regular deposition line is opened.

\section{Results and Discussion}\label{sec:RnD_OHOD}

\begin{figure}
\begin{center}
\includegraphics[width=85mm]{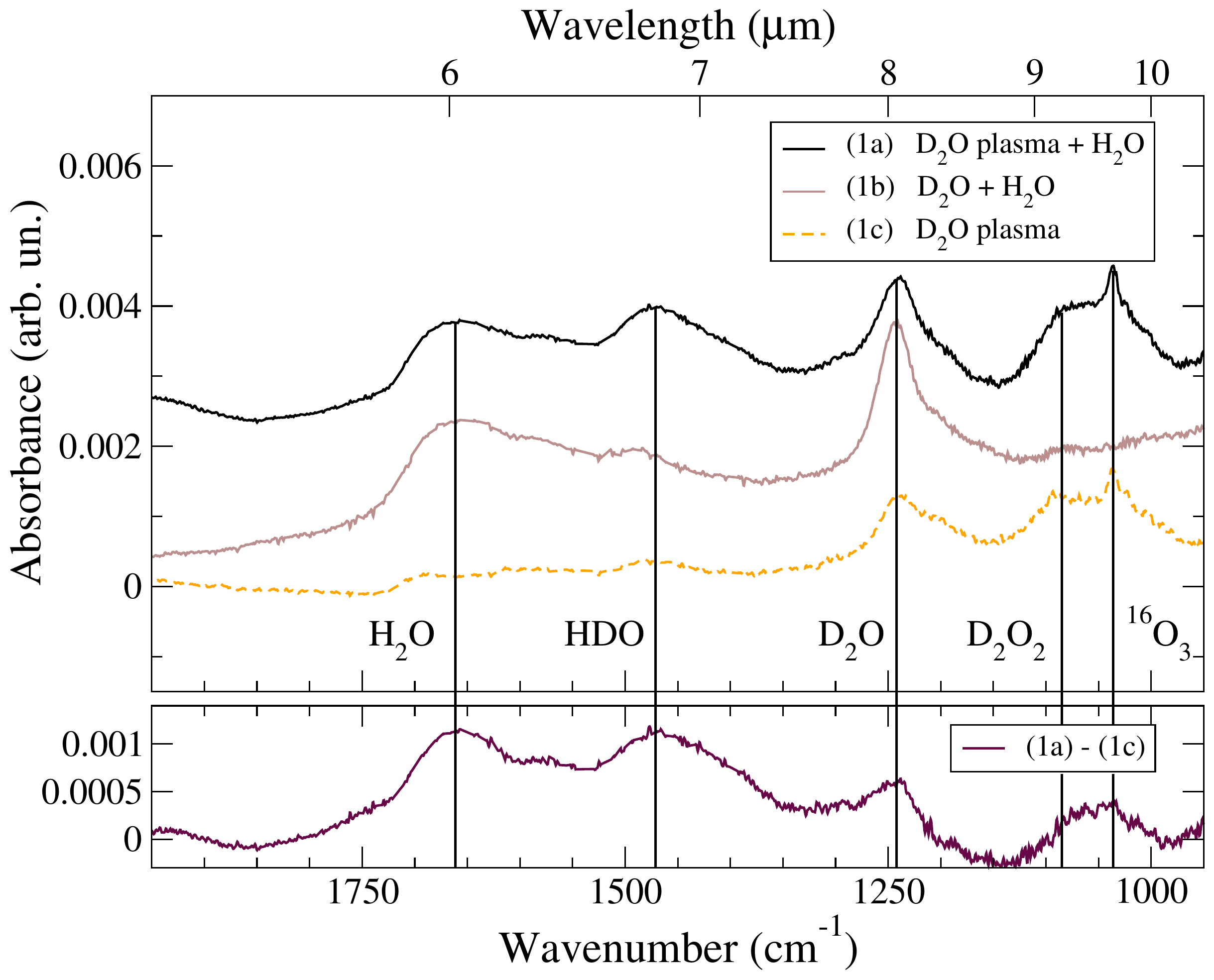}
\end{center}
\caption{\ce{H2O + OD}. RAIR spectra acquired after 90 minutes of co-deposition for experiments (1a)-(1c), Table~\ref{tab:exps_OHOD}. Graphs are offset for reasons of clarity. The lower panel shows a difference spectrum between experiments a and c (Table~\ref{tab:exps_OHOD}).}
\label{fig_H2O_OD}
\end{figure} 

\begin{figure}
\begin{center}
\includegraphics[width=85mm]{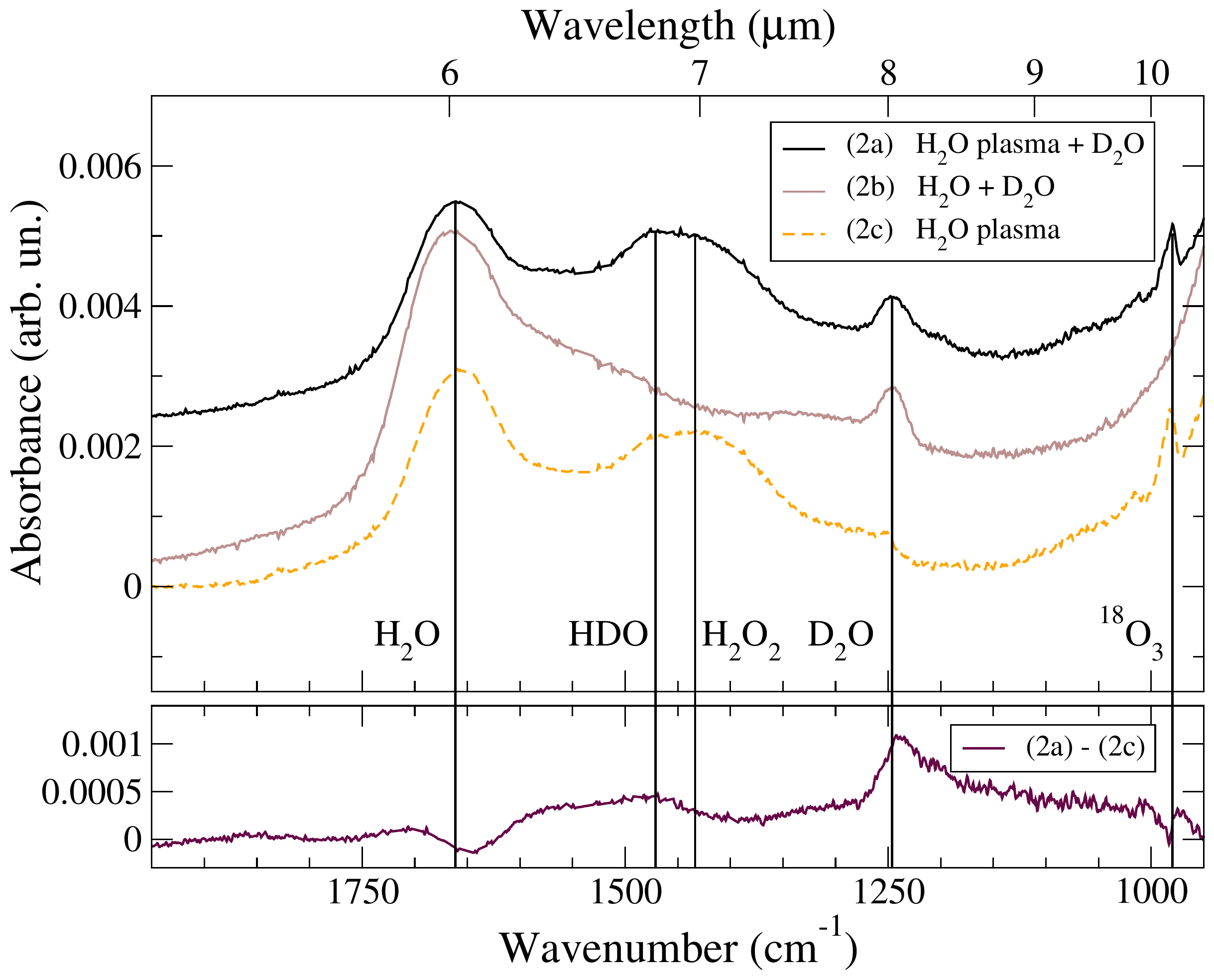}
\end{center}
\caption{\ce{D2O + OH}. RAIR spectra acquired after 90 minutes of co-deposition for experiments (2a)-(2c), Table~\ref{tab:exps_OHOD}. Graphs are offset for reasons of clarity. The lower panel shows a difference spectrum between experiments a and c (Table~\ref{tab:exps_OHOD}).}
\label{fig_D2O_OH}
\end{figure}

In Figures~\ref{fig_H2O_OD} and ~\ref{fig_D2O_OH}, the RAIR spectra recorded after 90 minutes of co-deposition are depicted for experiments 1a - 1c (to investigate reaction~\ref{R1}) and 2a - 2c (idem reaction~\ref{R4}) mentioned in Table~\ref{tab:exps_OHOD}. {The graphs represent nearly unprocessed data with only a two-point baseline subtracted (1900 and 4000 cm$^{-1}$). In both figures, a difference curve between experiment a and control experiment c is depicted as well. The overlapping peaks prohibit that direct quantitative information can be derived.} An overview of the products found in the various spectra and their respective origins is given in Table~\ref{exp_products}. A clear finding is that HDO is only seen for the reaction \ce{H2O + OD} (Figure~\ref{fig_H2O_OD}) and not for \ce{D2O + OH} (Figure~\ref{fig_D2O_OH}), indicative for different reaction efficiencies. {The interpretation of this observation is dicussed below.}

In Figure~\ref{fig_H2O_OD}, five characteristic infrared features are visible, namely for \ce{H2O}, \ce{HDO}, \ce{D2O}, \ce{D2O2}, and \ce{^{16}O3}. The presence of these species can be explained either by deposition, by (re-)combination of discharge products, or through additional reactions. During experiment 1a (\ce{H2O + OD}), three molecules are formed in the course of the co-deposition: \ce{HDO}, \ce{D2O2}, and \ce{O3}. Both deuterated peroxide and ozone are also visible in experiment 1c when only dissociated \ce{D2O} is deposited on the surface. This indicates that reactions between discharge fragments are responsible for the production of these species. The formation of \ce{HDO} is, however, observed as a result of reaction~\ref{R1} which only takes place when the plasma is switched on. The small amount of HDO observed in control experiment 1b, is attributed to contamination in the \ce{D2O} sample and is clearly much weaker compared to the feature present in experiment 1a. 

{To rule out that non-thermalized reaction mechanisms are at play here, such as the Eley-Rideal or hot-atom mechanism, an extra experiment was performed similar to experiment 1, but at a surface temperature of 60~K. At this temperature, the residence time of thermalized OD radicals on the surface is much shorter \citep{Oba:2011} and therefore if HDO formation is observed, this must proceed through a non-thermalized mechanism. However, the characteristic HDO peak was not detected throughout this experiment, which indicates that reaction~\ref{R1} on the surface takes place via the thermalized Langmuir-Hinshelwood mechanism.}

\begin{figure}
\begin{center}
\includegraphics[width=81mm]{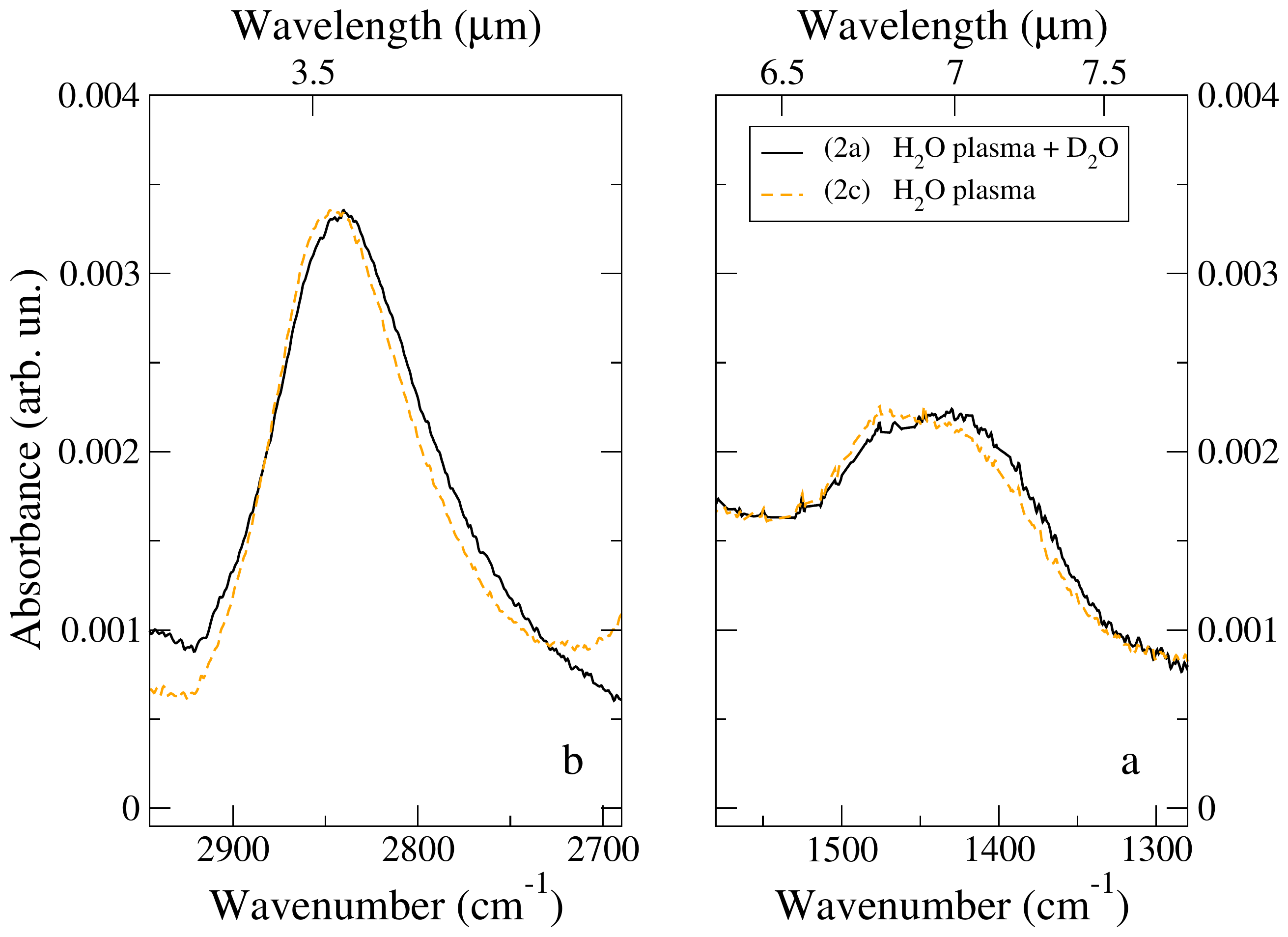}
\end{center}
\caption{Comparison of \ce{H2O2} RAIR spectra acquired after 90 minutes of co-deposition for experiments (2a) and (2c), Table~\ref{tab:exps_OHOD}. The right panel shows the OH bending wavelength domain studied here. The left panel is an additional spectrum in the $\nu_2 + \nu_6$ combination band domain.  Graphs are offset in such a way that the bands in both experiments overlap regardless of the baseline.}
\label{fig_comp_H2O2}
\end{figure}

In Figure~\ref{fig_D2O_OH}, four characteristic infrared features are observed, corresponding to \ce{H2O}, \ce{H2O2}, \ce{D2O}, and \ce{^{18}O3}. Here, the ozone consists of \ce{^{18}O} atoms, as a result of the \ce{H2^{18}O} plasma that produces both \ce{^{18}O2} and \ce{^{18}O}. Again, ozone and hydrogen peroxide are generated as a result of discharge fragment (re-)combinations.  Throughout experiment 2a (\ce{D2O + OH}), these are the only two molecules that are formed while HDO production seems to be lacking. {This also becomes clear when comparing the lower panels in Figures~\ref{fig_H2O_OD} and~\ref{fig_D2O_OH}. In Figure 2, the HDO peak around 1450 cm$^{-1}$ is much weaker.} However, the peak positions of hydrogen peroxide and singly deuterated water in water-rich environments are very close to each other: $\sim$1435 versus $\sim$1475 cm$^{-1}$ \citep{Devlin:1990, Oba:2014}. 
{Therefore, special care is needed to interpret this experiment and to conclude that HDO is not formed efficiently, as its absorption feature may be buried in the \ce{H2O2} signal. For a detectable level, one expects to see a clear shift and a change in total integrated intensity of the peak. In Figure~\ref{fig_comp_H2O2} a direct comparison between experiment 2a and control experiment 2c is shown. The right panel zooms in on the region of the OH bending mode of both HDO and \ce{H2O2} (discussed here), while the left panel zooms in on the $\nu_2 + \nu_6$ combination band of \ce{H2O2} at 2860 cm$^{-1}$. The latter is the only \ce{H2O2} band which does not overlap with other species \citep{Lannon:1971} and, as such, can be used as a reference point. }

{The figure shows that the spectra shift. The OH bending mode band shifts by about 20 cm$^{-1}$, halfway the peak positions of the \ce{H2O2} and HDO bands. At a first glance, this would be consistent with HDO formation, but a closer look learns that this shift is due to the changing matrix environment, \emph{i.e.}, the presence of \ce{D2O} in experiment 2a or lack thereof in experiment 2c, rather than a detectable amount of HDO formation. The first argument to support this is that also the combination band (at 2860 cm$^{-1}$) shifts, roughly 7 cm$^{-1}$, but in this case the band does not overlap with HDO and is therefore expected not to shift unless the matrix plays a role. }

{Secondly, to further study the influence of a mixed HDO and \ce{H2O2} ice on the bandwidth of the OH bending mode, we artificially added an HDO component to the \ce{H2O2} component of the \ce{H2O} plasma deposition. This is done by selecting specifically the HDO band in experiment 1a in Fig.~\ref{fig_H2O_OD}, setting its baseline to zero around 1350 and 1530 cm$^{-1}$, and adding it to the same region of the spectrum obtained in experiment 2c. This shows that the final band should then exhibit a larger bandwidth. Clearly the observed band in Fig.~\ref{fig_comp_H2O2} seems to have shifted rather than to have changed its profile because of merging features. This is only possible when HDO is not formed, \emph{i.e.}, the reaction \ce{D2O + OH -> OD + HDO} is not efficient. }

{The third argument follows from a comparison of the integrated band areas ($A$) of the OH bending mode in experiments 1a, 2a, and 2c. The ratio between ($A_\text{exp. 2a} - A_\text{exp. 2c}$) and $A_\text{exp. 1a}$ is derived
by integrating the three spectra over the same range and comparing the values. Since this involves a choice of the range over which to integrate, the three experiments were all integrated over several ranges. The  first integration point was chosen between 1340 and 1365 cm$^{-1}$ and the second point between 1520 and 1550 cm$^{-1}$. The values found for the ratio $k_4/k_1$ were all lower than 0.2, depending on the exact range chosen. Here, we remain conservative and stay with the upper limit and therefore set $k_4/k_1$ to 0.2. This factor is in agreement with the difference curve depicted in Fig.~\ref{fig_D2O_OH}. We therefore conclude that the observed shift in Fig.~\ref{fig_comp_H2O2} must be largely due to a matrix effect, ruling out an effective HDO formation in  reaction~\ref{R4}.}

\begin{table}
 \caption{List of assigned species in the experiments and their respective origins.}\label{exp_products}
\begin{tabular}{ll}
 \hline
\multicolumn{2}{l}{Experiment 1}\\
\hline
Products & Origin \\
\hline
\ce{H2O} & Deposition \\
\ce{HDO} & Reaction of \ce{H2O} with discharge fragments: \\
		&  \ce{H2O + OD -> OH + HDO} \\
	 &  \ce{OH + D -> HDO}\\
\ce{D2O} & Undissociated molecules in the plasma line \\
\ce{D2O2} & Reaction of discharge fragments: \\
	  & \ce{OD + OD -> D2O2} \\
	  & \ce{O2 + D + D -> D2O2}\\
\ce{O3} &  Reaction of discharge fragments: \\
	     & \ce{O + O -> O2} and \ce{O + O2 -> O3} \\
\hline
\multicolumn{2}{l}{Experiment 2}\\
\hline
Products & Origin \\
\hline
\ce{H2O}  & Undissociated molecules in the plasma line \\
\ce{H2O2}  &  Reaction of discharge fragments: \\
		& \ce{ OH + OH -> H2O2}\\
		& \ce{ O2 + H + H -> H2O2}\\
\ce{D2O}  & Deposition \\
\ce{O3}  & Reaction of discharge fragments: \\
	    &  \ce{O + O -> O2} and \ce{O + O2 -> O3} \\
\hline
\end{tabular}
\end{table}

A final, independent argument comes from theory. From the experimental set presented here, we conclude that reaction~\ref{R1} does, and reaction~\ref{R4} does not efficiently take place at a detectable level for the specific conditions studied. {To understand this difference and the overall likelihood of the reactions to take place at low temperatures, the gas-phase Gibbs free energies at 15~K are calculated using the enthalpies and entropies given by \citet{Chase:1998}:\\
\begin{tabular}{cr@{ }r@{ }l}
 \ref{R1}: & -1.2 kJ mol$^{-1}$ & = & -146~K\\
 \ref{R2}: & -1.3 kJ mol$^{-1}$ & = & -159~K\\
 \ref{R3}: & 1.2 kJ mol$^{-1}$ & = & \;146~K\\
 \ref{R4}: & 1.3 kJ mol$^{-1}$ & = & \;159~K.\\
\end{tabular}\\
The experimentally found difference in reaction probability for reactions~\ref{R1} and~\ref{R4} is in line with the predictions from gas-phase Gibbs free energies. Extrapolating the experiments, a similar behavior for reactions~\ref{R2} and~\ref{R3} is expected.

Additionally, from a tunneling point-of-view, the first two reactions transfer a hydrogen atom, whereas the final two reactions transfer a deuteron. Calculating the effective mass, $\mu$, for this system as outlined by \citet{Oba:2012} to find the kinetic isotope effect involved, we find a $\sim$25\% decrease in the reaction probabilities assuming that the tunneling rate is well-described by the Wentzel-Kramers-Brillouin or square-potential with width $a$ and height $E$ approximation: 
\begin{equation}
 k_\text{tunneling} \propto \exp \left( -\frac{2a}{h} \sqrt{2\mu E} \right) \;.
\end{equation}
This is in agreement with our experimental comparison between the first and the last reaction. Note, however, that the endergonicity still plays a role when tunneling is considered  
to further decrease the reaction probability \citep{Lamberts:2014B}.

To summarize all experimental and theoretical findings and considerations, we expect the following relations to hold: 
\begin{align*}
 k_4 &\ll k_1 \mbox{ (with an absolute upper limit } k_4 < 0.2\;k_1) \\
k_1 &\approx k_2 \mbox{ (assuming thermal activation only)} \\
k_1 &> k_2 \mbox{ (assuming that tunneling plays a role)} \\
k_3 &\approx k_4\;.
\end{align*}
{We cannot draw any definitive conclusions concerning reactions~\ref{Rlast} and~\ref{Rlast2}, although it is important to note that the change in Gibbs free is zero on average and if tunneling is involved $k_6 < k_5 $. }

\begin{table}
 \caption{{Parameters used in the astrochemical simulations for a translucent cloud. See text regarding the values for $n(\text{O(I)})$, $n(\text{HD})$, $n(\text{\ce{H2}})$ and $n(\text{D})$.}}\label{tab:cloud_par}
\begin{tabular}{cccccc}
\hline
 $n(\text{\ce{H2}})$ & $n(\text{HD})$ &  $n(\text{H(I)})$ & $n(\text{D(I)})$ & $n(\text{O(I)})$ \\
cm$^{-3}$ &  cm$^{-3}$  &  cm$^{-3}$ &  cm$^{-3}$ &  cm$^{-3}$   \\
$5 \times 10^2$  & $8\times 10^{-4}$  & 2  & $6 \times 10^{-3}$  & $3\times 10^{-1}$  \\
\hline
\hline
A$_V$ & $T_\text{gas}$ & $T_\text{grain}$ & $T_\text{exc.}$ & & \\
mag & K & K & K & & \\
3 & 14 & 20 & 700 & & \\
\hline
\end{tabular}
\end{table}

\section{Astrochemical Implications}\label{sec:AC_OHOD}

The key to the astrochemical relevance of the reactions studied here lies in the relative abundances of the water isotopes ($n(\ce{H2O}) > n(\ce{HDO}) > n(\ce{D2O})$) and the ice abundance of both hydroxyl radical isotopes, OH and OD. These species can be either formed on the interstellar dust surfaces \emph{via} \ce{O + H} and \ce{O + D} reactions or \emph{via} photodissociation of frozen \ce{H2O}, \ce{HDO} or \ce{D2O}. Therefore, especially in rather low-A$_\text{V}$ regions such as translucent clouds, the deuterium enrichment effect of the reactions studied here can be substantial. In these regions, water is continuously formed and destroyed on the surface until a steady state is reached \citep{Cuppen:2007A, Cazaux:2010, Lamberts:2014}. Each time a formed HDO molecule is dissociated into \ce{OH + D} or, more likely, \ce{OD + H} \citep{Koning:2013}, there is a chance that the hydroxyl radical evaporates. The OD evaporation is in competition with three solid-state reactions creating or recreating HDO: 
\begin{align}
\ce{H + OD} &\rightarrow \ce{HDO}  \;\; \tag{R7}\label{R5} \\
 \ce{H2 + OD} &\rightarrow \ce{H + HDO} \;\; \tag{R8}\label{R6} \\
\ce{H2O + OD} &\rightarrow \ce{OH + HDO} \;\;\tag{R1} 
\end{align}

{To further characterize this, a previously-used continuous-time random-walk Kinetic Monte Carlo model -- including desorption, diffusion, reaction, and evaporation processes -- for water formation \citep{Lamberts:2013, Lamberts:2014} is adapted here with an extension of the chemical surface network including all deuteration reactions. }
All hydrogenation reactions are duplicated and replaced with deuterium analogue(s). {Deuterium chemistry is} initiated by surface reactions with HD or D. If tunneling is expected to be involved, the activation energies are altered accordingly \citep{Oba:2012, Oba:2014}. Furthermore, for reactions that can result in more than one product, the branching ratios between the product channels are distributed statistically. We realize that this model is a rather crude approximation; {the intention is not to derive accurate values since the number of involved reactions and dependencies is too extensive for this.} It merely serves to test which parameters are crucial when including OH and OD abstraction reactions in the solid state water formation network.

{The parameters of the `translucent cloud' studied here are summarized in Table~\ref{tab:cloud_par}}. They are chosen such to have a high enough photon flux to induce photodissociation of water, but simulatenously to allow for a minimal build-up of an icy layer. A high D abundance (one order of magnitude higher compared to Figure~4 in \citep{Lepetit:2002} for a molecular fraction of $\sim 0.9$) is adapted to enhance the HDO abundance on the surface to reduce the total simulation time. The atomic hydrogen density is chosen to be 2 cm$^{-3}$ \citep{Goldsmith:2005}. The remaining fractional densities of the species involved have been chosen typical for a cloud with $n_\text{H}$ = 1000 cm$^{-3}$, \emph{i.e.}, $n(\text{HD})/n_\text{H} = 8\times 10^{-7}$ and $n(\text{O(I)})/n_\text{H} = 3\times 10^{-4}$ \citep{Lepetit:2002, Nguyen:2002}. {The binding energies of the deuterated species are copied from their regular counterparts, \emph{e.g.}, the binding energy of OH and OD radicals on a flat surface is $\sim$650~K, but increases with 210~K for each local neighbor as explained in \citet{Lamberts:2013}. This binding energy is quite low compared to recent experimental studies \citep{He:2014} and therefore we have performed additional simulation runs doubling the binding energy.}

\begin{table}
 \caption{{Specification of the parameters that have been studied specificially in this paper, $E_\text{bind, OH}$, $k_1$, $k_2$, and $k_3$, $k_4$ for the six runs performed with the resulting HDO/\ce{H2O} ice ratio. The ratios are averaged over five simulation runs each.}}\label{tab:cloud_par2}
 \begin{tabular}{c@{\;\;\;}c@{\;\;\;}c@{\;\;}c@{\;\;}c@{\;\;	}c}
 \hline
  & $k_1$, $k_2$ & $k_3$, $k_4$  & $k_5$, $k_6$ & $E_\text{b, OH}$	& HDO/\ce{H2O}\\
  & s$^{-1}$	 & s$^{-1}$	 & s$^{-1}$	& K 	& ($\times  10^{-3}$)\\
  \hline
I & --			&  --			&   $1.9 \times 10^{-10}$	& 650 	& 3.0   \\
II& $1.1 \times 10^5$	&  $2.2 \times 10^4$ 	&   $1.9 \times 10^{-10}$	& 650 	& 6.1   \\
III& $1.1 \times 10^5$	&  -- 			&   $1.9 \times 10^{-10}$	& 650 	& 8.4   \\
IV & --			&  --			&   $1.9 \times 10^{-10}$	& 1300 	& 4.3   \\
V& $1.1 \times 10^5$	&  $2.2 \times 10^4$ 	&   $1.9 \times 10^{-10}$	& 1300 	& 7.4   \\
  \hline
 \end{tabular}
\end{table}

{All simulations have been performed five times to increase the low S/N ratio that is inherent to the low HDO abundances. The results for different runs are summarized in Table~\ref{tab:cloud_par2} and visualized in Fig.~\ref{fig_abun}. The HDO/\ce{H2O} values mentioned in Table~\ref{tab:cloud_par2} are average values with standard deviations of approximately 35\% corresponding to a time of $6.5 \times 10^4$ years.}

{Run I neglects the effects studied in this work, but includes reactions~\ref{Rlast} and~\ref{Rlast2}; runs II and III incorporate~\ref{R1}-\ref{Rlast2} and ~\ref{R1},~\ref{R2},~\ref{Rlast}, and \ref{Rlast2}, respectively, for a binding energy of 650~K. Run IV is similar to run I, but for 1300~K. Run V is the 1300~K equivalent of run II. The relations for $k_1-k_4$ derived experimentally are implemented in the Monte Carlo routine in two ways. Firstly, all reactions are incorporated with a non-zero rate, assuming a conservative upper limit corresponding to  $(k_3,\;k_4) < 0.2\;(k_1,\; k_2)$, {\emph{i.e.}, simulation run II}. Secondly, as experimentally no clear evidence of the occurence of reaction~\ref{R4} is found, both $k_3$ and $k_4$ are set to zero in {run III}. These simulations are compared to run {I}, where reactions~\ref{R1}-\ref{Rlast2} are excluded from the network. 
The reaction rates of the reactions~\ref{Rlast} and~\ref{Rlast2} are equal throughout all simulations, but have been reduced artificially to reduce the computational cost of the simulation and avoid that a series of reactions between two neighboring species keeps occuring without a net change in OH and \ce{H2O} abundance.
}

\begin{figure}
\begin{center}
\includegraphics[width=85mm]{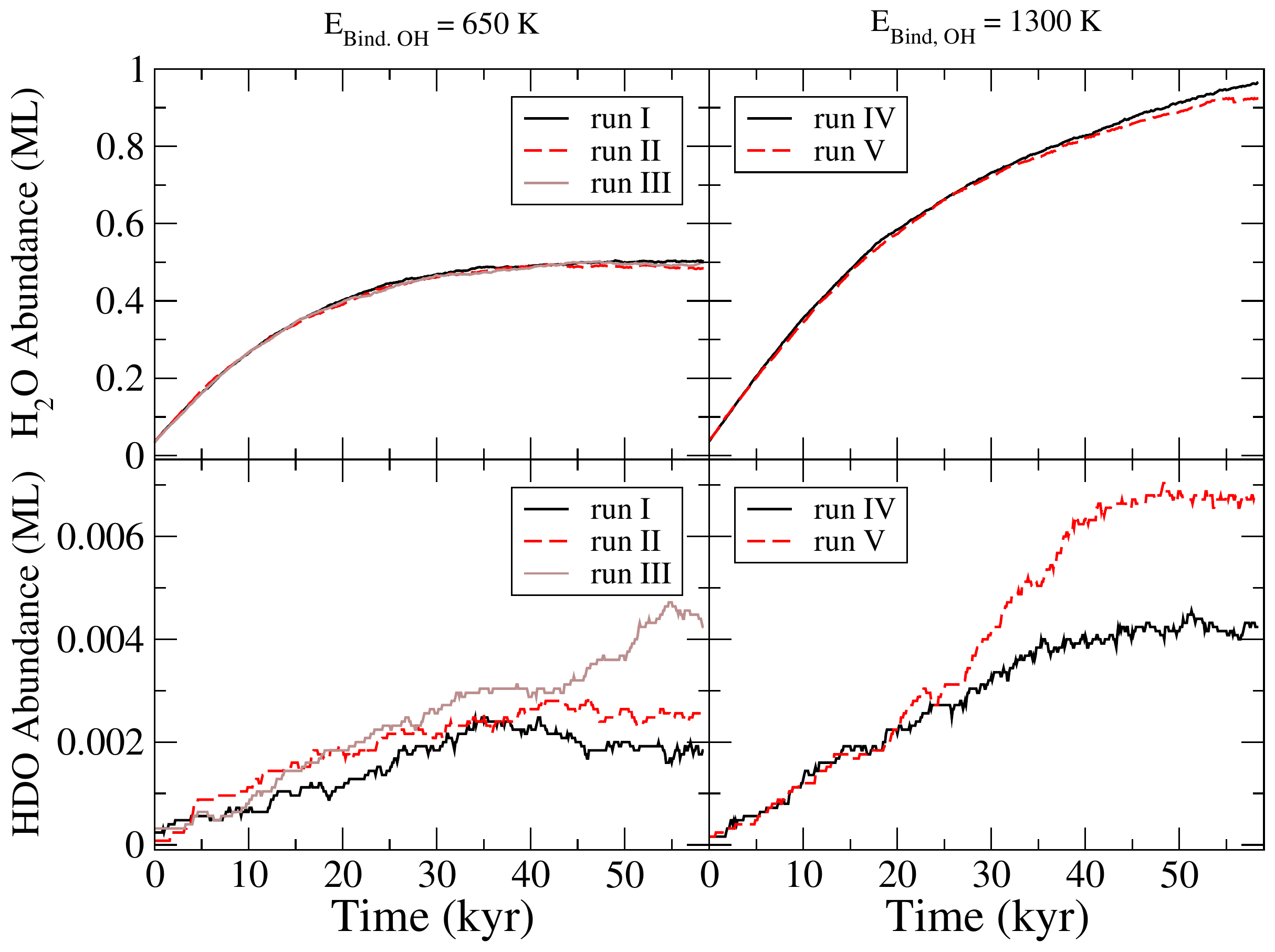}
\end{center}
\caption{Time evolution of the average \ce{H2O} and HDO abundances for simulation runs I-V. }
\label{fig_abun}
\end{figure}

{The influence of including reactions~\ref{R1}-\ref{Rlast2} in the network as well as of different OH and OD binding energies on the ice HDO/\ce{H2O} ratio is discussed below. It is directly clear that there is no substantial effect on the \ce{H2O} formation (upper two panels). In the case of HDO, it takes time before the reactions discussed here start changing abundances w.r.t. a model in which these are not taken into account (run I). Reactions~\ref{R1} and~\ref{R3} start to play a role only after sufficient HDO has been produced on the surface}, \emph{i.e.}, either at later stages in the simulation ($\;>3\times 10^4$ years) or upon artificial increase of the D abundance. Indeed, reactions~\ref{R2},~\ref{R4}, and~\ref{Rlast2} do not take place in the simulations, as a result of the low concentrations of OD and \ce{D2O}.

We find that the {absolute HDO abundance, and therefore also the HDO/\ce{H2O} ice ratio increases upon including the studied reactions in the reaction network, compare simulation I to II/III, and simulation IV to V. Each time that an OD radical is created -- either by reaction or dissociation of HDO -- this species can easily regenerate deuterated water. }

{The same conclusions hold upon an increase in the binding energy, from 650 to 1300~K on a flat surface, of the OH and OD radicals. This primarily results in thicker ice layers, since less desorption takes place, which can be seen directly from Fig.~\ref{fig_abun}.}

{Finally, including reactions~\ref{R1}-\ref{Rlast2} also decreases the absolute abundance of deuterated peroxide, \ce{HDO2}, and the ratio \ce{HDO2}/\ce{H2O2} by a factor 1.5 to 4.5, depending on the simulation run. An OD radical is more likely to react with \ce{H2O} than to find a OH fragment and form \ce{HDO2}.}

{In summary: incorporating reactions~\ref{R1}-\ref{R4} can change the \ce{HDO}/\ce{H2O} ratio by as much as a factor 2-3 compared to models not taking into account these reactions.}

\section{Conclusions}\label{sec:concl_OHOD}

We demonstrated experimentally that the reaction between water and a deuterated hydroxyl radical produces singly deuterated water, and that the deuterated analogue of this reaction does not proceed with the same efficiency, namely:
\begin{align*}
& \ce{H2O + OD} \xrightarrow{k_1} \ce{OH + HDO} \;\; \tag{R1}\\
& \ce{D2O + OH} \xrightarrow{k_4} \ce{OD + HDO} \;\; \tag{R4}\\
\end{align*}
with $k_4 < 0.2\;k_1 $.
This result is theoretically supported by the difference in gas-phase Gibbs free energy of the reactions which renders the reaction~\ref{R1} exergonic, but the reaction~\ref{R4} endergonic. If such a proton/deuteron transfer reaction was to occur (partially) \emph{via} tunneling, the kinetic isotope effect also points towards a lower reaction rate for the deuteron transfer, \emph{i.e.}, $k_4 < k_1$.

Furthermore, a deuterated-water-formation Kinetic Monte Carlo model shows that including reactions between water and hydroxyl radical isotopes can change the HDO/\ce{H2O} ice ratio with respect to surface formation schemes neglecting these processes. The extent of this change depends on the exact parameters included as well as on the astronomical conditions. More information on the exact role of crucial parameters, such as precise values for the reaction rates and the role of the excess heat upon reaction, in general, is needed before this model can be further extended. 

Placing this in context with other studies, we want to stress that the previous experimental solid-state reactions involving different reaction efficiencies for different isotopologues all point towards the involvement of tunneling in overcoming the reaction activation barrier \citep{Oba:2012, Oba:2014}. This essentially results in H-enrichment of the ices since tunneling favors reactions with hydrogen over those with deuterium. The reactions discussed in the present work, however, could offer a pathway to D-enrichment. In light of the recent investigations of the \ce{HDO}/\ce{H2O} and \ce{D2O}/\ce{H2O} ratios, it is thus pivotal to understand the isotopic effect of all single reactions included in the water formation network \citep{Dishoeck:2013}. The present study adds one specific reaction channel that is relevant within the large picture. {Although it is often implicitly assumed that the ratios in the ice and in the gas are coupled, this is not always the case \citep{Taquet:2014, Furuya:inprep}.}

Finally, the importance of the abstraction reactions induced by the hydroxyl radical previously reported \citep{Oba:2012, Garrod:2013} is reinforced by the present results, especially since this radical plays an important role as reactive intermediate in water formation. Such reactions can strongly affect the local OH (and OD) abundance in interstellar ices, not only \emph{via} abstraction reactions, but also influence, \emph{e.g.}, \ce{CO2} formation, \emph{via} the diffusion analog reaction~\ref{Rlast} (\ce{H2O + OH -> H2O + OH}), in the case that \ce{CO2} forms through \ce{CO + OH} recombination \citep{Mennella:2004, Oba:2011, Ioppolo:2011b, Noble:2011}.

\section{Acknowledgements}
We thank Magnus Persson, Kenji Furuya, and Xander Tielens for stimulating discussions. 

Astrochemistry in Leiden in general and the visit of F.P. to our laboratory in particular was supported by the European Community’s Seventh Framework Programme (FP7/2007- 2013) under grant agreement n.238258, the Netherlands Research School for Astronomy (NOVA) and from the Netherlands Organization for Scientific Research (NWO) through a VICI grant. T.L. is supported by the Dutch Astrochemistry Network financed by The Netherlands Organization for Scientific Research (NWO). Support for S.I. from the Marie Curie Fellowship (FP7-PEOPLE-2011-IOF-300957) is gratefully acknowledged. H.M.C. is grateful for support from the VIDI research program 700.10.427, which is financed by The Netherlands Organization for Scientific Research (NWO) and from the European Research Council (ERC-2010-StG, Grant Agreement no. 259510-KISMOL).

\bibliographystyle{mn2e}
\bibliography{biblio_OHOD}

\begin{thebibliography}{}

\bibitem[\protect\citeauthoryear{{Andersson} \& {van Dishoeck}}{{Andersson} \&
  {van Dishoeck}}{2008}]{Andersson:2008}
{Andersson} S.,  {van Dishoeck} E.~F.,  2008, \aap, 491, 907

\bibitem[\protect\citeauthoryear{Anton, Wiegner, Naumann, Liebmann, Klein \&
  Bradley}{Anton et~al.}{2000}]{Anton:2000}
Anton R.,  Wiegner T.,  Naumann W.,  Liebmann M.,  Klein C.,    Bradley C.,
  2000, Rev.~Sci.~Instrum.~, 71, 1177

\bibitem[\protect\citeauthoryear{Arasa, van Hemert, van Dishoeck \&
  Kroes}{Arasa et~al.}{2013}]{Arasa:2013}
Arasa C.,  van Hemert M.~C.,  van Dishoeck E.~F.,    Kroes G.~J.,  2013, The
  Journal of Physical Chemistry A, 117, 7064

\bibitem[\protect\citeauthoryear{{Caselli} \& {Ceccarelli}}{{Caselli} \&
  {Ceccarelli}}{2012}]{Caselli:2012}
{Caselli} P.,  {Ceccarelli} C.,  2012, \aapr, 20, 56

\bibitem[\protect\citeauthoryear{Cazaux, Cobut, Marseille, Spaans \&
  Caselli}{Cazaux et~al.}{2010}]{Cazaux:2010}
Cazaux S.,  Cobut V.,  Marseille M.,  Spaans M.,    Caselli P.,  2010,
  Astron.~Astrophys., 522, A74

\bibitem[\protect\citeauthoryear{{Chaabouni}, {Minissale}, {Manic{\`o}},
  {Congiu}, {Noble}, {Baouche}, {Accolla}, {Lemaire}, {Pirronello} \&
  {Dulieu}}{{Chaabouni} et~al.}{2012}]{Chaabouni:2012a}
{Chaabouni} H.,  {Minissale} M.,  {Manic{\`o}} G.,  {Congiu} E.,  {Noble}
  J.~A.,  {Baouche} S.,  {Accolla} M.,  {Lemaire} J.~L.,  {Pirronello} V.,
  {Dulieu} F.,  2012, \jcp, 137, 234706

\bibitem[\protect\citeauthoryear{{Chang} \& {Herbst}}{{Chang} \&
  {Herbst}}{2014}]{Chang:2014}
{Chang} Q.,  {Herbst} E.,  2014, \apj, 787, 135

\bibitem[\protect\citeauthoryear{Chase}{Chase}{1998}]{Chase:1998}
Chase M.~W.,  1998, Journal of physical and chemical reference data, Monograph,
  no. 9..
American Chemical Society, Washington DC

\bibitem[\protect\citeauthoryear{Cheng, Lee \& Ogilvie}{Cheng
  et~al.}{1988}]{Cheng:1988}
Cheng B.-M.,  Lee Y.-P.,    Ogilvie J.,  1988, Chemical Physics Letters, 151,
  109

\bibitem[\protect\citeauthoryear{{Cuppen} \& {Herbst}}{{Cuppen} \&
  {Herbst}}{2007}]{Cuppen:2007A}
{Cuppen} H.~M.,  {Herbst} E.,  2007, Astrophys.~ J., 668, 294

\bibitem[\protect\citeauthoryear{{Cuppen}, {Ioppolo}, {Romanzin} \&
  {Linnartz}}{{Cuppen} et~al.}{2010}]{Cuppen:2010B}
{Cuppen} H.~M.,  {Ioppolo} S.,  {Romanzin} C.,    {Linnartz} H.,  2010, \pccp,
  12, 12077

\bibitem[\protect\citeauthoryear{Devlin}{Devlin}{1990}]{Devlin:1990}
Devlin J.,  1990, Journal of Molecular Structure, 224, 33

\bibitem[\protect\citeauthoryear{Do N.~H.}{Do}{2014}]{Do:2014}
Do N.~H. V.~D. C. P.~D.,  2014, \mnras, 443, 207

\bibitem[\protect\citeauthoryear{{Dulieu}}{{Dulieu}}{2011}]{Dulieu:2011IAU}
{Dulieu} F.,  2011, in {Cernicharo} J.,  {Bachiller} R.,  eds, IAU Symposium
  Vol.~280 of IAU Symposium, {Water Ice Formation and the o/p Ratio}.
pp 405--415

\bibitem[\protect\citeauthoryear{Furuya}{Furuya}{2015}]{Furuya:inprep}
Furuya K.,  2015, Personal communication, Article in preparation

\bibitem[\protect\citeauthoryear{Garrod}{Garrod}{2013}]{Garrod:2013}
Garrod R.~T.,  2013, The Astrophysical Journal, 765, 60

\bibitem[\protect\citeauthoryear{{Goldsmith} \& {Li}}{{Goldsmith} \&
  {Li}}{2005}]{Goldsmith:2005}
{Goldsmith} P.~F.,  {Li} D.,  2005, \apj, 622, 938

\bibitem[\protect\citeauthoryear{{Hartogh}, {Lis}, {Bockel{\'e}e-Morvan}, {de
  Val-Borro}, {Biver}, {K{\"u}ppers}, {Emprechtinger}, {Bergin}, {Crovisier},
  {Rengel}, {Moreno}, {Szutowicz} \& {Blake}}{{Hartogh}
  et~al.}{2011}]{Hartogh:2011}
{Hartogh} P.,  {Lis} D.~C.,  {Bockel{\'e}e-Morvan} D.,  {de Val-Borro} M.,
  {Biver} N.,  {K{\"u}ppers} M.,  {Emprechtinger} M.,  {Bergin} E.~A.,
  {Crovisier} J.,  {Rengel} M.,  {Moreno} R.,  {Szutowicz} S.,    {Blake}
  G.~A.,  2011, \nat, 478, 218

\bibitem[\protect\citeauthoryear{{He} \& {Vidali}}{{He} \&
  {Vidali}}{2014}]{He:2014}
{He} J.,  {Vidali} G.,  2014, \apj, 788, 50

\bibitem[\protect\citeauthoryear{Hiraoka, Miyagoshi, Takayama, Yamamoto \&
  Kihara}{Hiraoka et~al.}{1998}]{Hiraoka:1998}
Hiraoka K.,  Miyagoshi T.,  Takayama T.,  Yamamoto K.,    Kihara Y.,  1998,
  Astrophys.~J.~, 498, 710

\bibitem[\protect\citeauthoryear{Ioppolo, Fedoseev, Lamberts, Romanzin \&
  Linnartz}{Ioppolo et~al.}{2013}]{Ioppolo:2013}
Ioppolo I.,  Fedoseev G.,  Lamberts T.,  Romanzin C.,    Linnartz H.,  2013,
  Rev.~Sci.~Instrum.~, 84, 073112

\bibitem[\protect\citeauthoryear{{Ioppolo}, {Cuppen}, {Romanzin}, {van
  Dishoeck} \& {Linnartz}}{{Ioppolo} et~al.}{2008}]{Ioppolo:2008}
{Ioppolo} S.,  {Cuppen} H.~M.,  {Romanzin} C.,  {van Dishoeck} E.~F.,
  {Linnartz} H.,  2008, Astrophys.~ J., 686, 1474

\bibitem[\protect\citeauthoryear{{Ioppolo}, {Cuppen}, {Romanzin}, {van
  Dishoeck} \& {Linnartz}}{{Ioppolo} et~al.}{2010}]{Ioppolo:2010}
{Ioppolo} S.,  {Cuppen} H.~M.,  {Romanzin} C.,  {van Dishoeck} E.~F.,
  {Linnartz} H.,  2010, \pccp, 12, 12065

\bibitem[\protect\citeauthoryear{{Ioppolo}, {van Boheemen}, {Cuppen}, {van
  Dishoeck} \& {Linnartz}}{{Ioppolo} et~al.}{2011}]{Ioppolo:2011b}
{Ioppolo} S.,  {van Boheemen} Y.,  {Cuppen} H.~M.,  {van Dishoeck} E.~F.,
  {Linnartz} H.,  2011, \mnras, 412

\bibitem[\protect\citeauthoryear{Itikawa \& Mason}{Itikawa \&
  Mason}{2005}]{Itikawa:2005}
Itikawa Y.,  Mason N.,  2005, Journal of Physical and Chemical Reference Data,
  34

\bibitem[\protect\citeauthoryear{Koning, Kroes \& Arasa}{Koning
  et~al.}{2013}]{Koning:2013}
Koning J.,  Kroes G.~J.,    Arasa C.,  2013, The Journal of Chemical Physics,
  138,

\bibitem[\protect\citeauthoryear{{Kristensen}, {Amiaud}, {Fillion}, {Dulieu} \&
  {Lemaire}}{{Kristensen} et~al.}{2011}]{Kristensen:2011}
{Kristensen} L.~E.,  {Amiaud} L.,  {Fillion} J.-H.,  {Dulieu} F.,    {Lemaire}
  J.-L.,  2011, \aap, 527, A44

\bibitem[\protect\citeauthoryear{{Lamberts}, {Cuppen}, {Fedoseev}, {Ioppolo},
  {Chuang} \& {Linnartz}}{{Lamberts} et~al.}{2014}]{Lamberts:2014B}
{Lamberts} T.,  {Cuppen} H.~M.,  {Fedoseev} G.,  {Ioppolo} S.,  {Chuang} K.-J.,
     {Linnartz} H.,  2014, \aap, 570, A57

\bibitem[\protect\citeauthoryear{Lamberts, Cuppen, Ioppolo \&
  Linnartz}{Lamberts et~al.}{2013}]{Lamberts:2013}
Lamberts T.,  Cuppen H.~M.,  Ioppolo I.,    Linnartz H.,  2013, \pccp, 15, 8287

\bibitem[\protect\citeauthoryear{Lamberts, de Vries \& Cuppen}{Lamberts
  et~al.}{2014}]{Lamberts:2014}
Lamberts T.,  de Vries X.,    Cuppen H.~M.,  2014, Far. Disc., 168, 327

\bibitem[\protect\citeauthoryear{Lamberts, Ioppolo, Cuppen, Fedoseev \&
  Linnartz}{Lamberts et~al.}{2015}]{Lamberts:2015}
Lamberts T.,  Ioppolo S.,  Cuppen H.~M.,  Fedoseev G.,    Linnartz H.,  2015,
  subm.

\bibitem[\protect\citeauthoryear{Lannon, Verderame \& Anderson}{Lannon
  et~al.}{1971}]{Lannon:1971}
Lannon J.~A.,  Verderame F.~D.,    Anderson R.~W.,  1971, The Journal of
  Chemical Physics, 54

\bibitem[\protect\citeauthoryear{{Le Petit}, {Roueff} \& {Le Bourlot}}{{Le
  Petit} et~al.}{2002}]{Lepetit:2002}
{Le Petit} F.,  {Roueff} E.,    {Le Bourlot} J.,  2002, \aap, 390, 369

\bibitem[\protect\citeauthoryear{{Lipshtat}, {Biham} \& {Herbst}}{{Lipshtat}
  et~al.}{2004}]{Lipshtat:2004}
{Lipshtat} A.,  {Biham} O.,    {Herbst} E.,  2004, \mnras, 348, 1055

\bibitem[\protect\citeauthoryear{Lloyd}{Lloyd}{1974}]{Lloyd:1974}
Lloyd A.~C.,  1974, International Journal of Chemical Kinetics, 6, 169

\bibitem[\protect\citeauthoryear{{Mennella}, {Palumbo} \& {Baratta}}{{Mennella}
  et~al.}{2004}]{Mennella:2004}
{Mennella} V.,  {Palumbo} M.~E.,    {Baratta} G.~A.,  2004, \apj, 615, 1073

\bibitem[\protect\citeauthoryear{Meyer \& Reuter}{Meyer \&
  Reuter}{2014}]{Meyer:2014}
Meyer J.,  Reuter K.,  2014, Angewandte Chemie International Edition, 53, 4721

\bibitem[\protect\citeauthoryear{Miyauchi, Hidaka, Chigai, Nagaoka, Watanabe \&
  Kouchi}{Miyauchi et~al.}{2008}]{Miyauchi:2008}
Miyauchi N.,  Hidaka H.,  Chigai T.,  Nagaoka A.,  Watanabe N.,    Kouchi A.,
  2008, Chem.~Phys.~Lett., 456, 27

\bibitem[\protect\citeauthoryear{{Nagaoka}, {Watanabe} \& {Kouchi}}{{Nagaoka}
  et~al.}{2005}]{Nagaoka:2005}
{Nagaoka} A.,  {Watanabe} N.,    {Kouchi} A.,  2005, \apjl, 624, L29

\bibitem[\protect\citeauthoryear{{Nagaoka}, {Watanabe} \& {Kouchi}}{{Nagaoka}
  et~al.}{2007}]{Nagaoka:2007}
{Nagaoka} A.,  {Watanabe} N.,    {Kouchi} A.,  2007, The Journal of Physical
  Chemistry A, 111, 3016

\bibitem[\protect\citeauthoryear{{Nguyen}, {Ruffle}, {Herbst} \&
  {Williams}}{{Nguyen} et~al.}{2002}]{Nguyen:2002}
{Nguyen} T.~K.,  {Ruffle} D.~P.,  {Herbst} E.,    {Williams} D.~A.,  2002,
  \mnras, 329, 301

\bibitem[\protect\citeauthoryear{Noble, Dulieu, Congiu \& Fraser}{Noble
  et~al.}{2011}]{Noble:2011}
Noble J.~A.,  Dulieu F.,  Congiu E.,    Fraser H.~J.,  2011, The Astrophysical
  Journal, 735, 121

\bibitem[\protect\citeauthoryear{Oba, Osaka, Watanabe, Chigai \& Kouchi}{Oba
  et~al.}{2014}]{Oba:2014}
Oba Y.,  Osaka K.,  Watanabe N.,  Chigai T.,    Kouchi A.,  2014, Far.~ Disc.~,
  168, 185

\bibitem[\protect\citeauthoryear{Oba, Watanabe, Hama, Huwahata, Hidaka \&
  Kouchi}{Oba et~al.}{2012}]{Oba:2012}
Oba Y.,  Watanabe N.,  Hama T.,  Huwahata K.,  Hidaka H.,    Kouchi A.,  2012,
  Astrophys.~J., 749, 67

\bibitem[\protect\citeauthoryear{Oba, Watanabe, Kouchi, Hama \& Pirronello}{Oba
  et~al.}{2011}]{Oba:2011}
Oba Y.,  Watanabe N.,  Kouchi A.,  Hama T.,    Pirronello V.,  2011, Phys.
  Chem. Chem. Phys., 13, 15792

\bibitem[\protect\citeauthoryear{{{\"O}berg}, Linnartz, Visser \& van
  Dishoeck}{{{\"O}berg} et~al.}{2009}]{Oberg:2009B}
{{\"O}berg} K.~I.,  Linnartz H.,  Visser R.,    van Dishoeck E.~F.,  2009,
  \apj, 693, 1209

\bibitem[\protect\citeauthoryear{{Oliveira}, {H{\'e}brard}, {Howk}, {Kruk},
  {Chayer} \& {Moos}}{{Oliveira} et~al.}{2003}]{Oliveira:2003}
{Oliveira} C.~M.,  {H{\'e}brard} G.,  {Howk} J.~C.,  {Kruk} J.~W.,  {Chayer}
  P.,    {Moos} H.~W.,  2003, \apj, 587, 235

\bibitem[\protect\citeauthoryear{{Piskunov}, {Wood}, {Linsky}, {Dempsey} \&
  {Ayres}}{{Piskunov} et~al.}{1997}]{Piskunov:1997}
{Piskunov} N.,  {Wood} B.~E.,  {Linsky} J.~L.,  {Dempsey} R.~C.,    {Ayres} R.,
   1997, \apj, 474, 315

\bibitem[\protect\citeauthoryear{{Roberts}, {Herbst} \& {Millar}}{{Roberts}
  et~al.}{2003}]{Roberts:2003}
{Roberts} H.,  {Herbst} E.,    {Millar} T.~J.,  2003, \apjl, 591, L41

\bibitem[\protect\citeauthoryear{{Rodgers} \& {Charnley}}{{Rodgers} \&
  {Charnley}}{2002}]{Rodgers:2002}
{Rodgers} S.~D.,  {Charnley} S.~B.,  2002, \mnras, 330, 660

\bibitem[\protect\citeauthoryear{{Romanzin}, {Ioppolo}, {Cuppen}, {van
  Dishoeck} \& {Linnartz}}{{Romanzin} et~al.}{2011}]{Romanzin:2011}
{Romanzin} C.,  {Ioppolo} S.,  {Cuppen} H.~M.,  {van Dishoeck} E.~F.,
  {Linnartz} H.,  2011, \jcp, 134, 084504

\bibitem[\protect\citeauthoryear{Straub, Lindsay, Smith \& Stebbings}{Straub
  et~al.}{1998}]{Straub:1998}
Straub H.~C.,  Lindsay B.~G.,  Smith K.~A.,    Stebbings R.~F.,  1998, The
  Journal of Chemical Physics, 108

\bibitem[\protect\citeauthoryear{{Taquet}, {Charnley} \& {Sipil{\"a}}}{{Taquet}
  et~al.}{2014}]{Taquet:2014}
{Taquet} V.,  {Charnley} S.~B.,    {Sipil{\"a}} O.,  2014, \apj, 791, 1

\bibitem[\protect\citeauthoryear{{Tielens}}{{Tielens}}{1983}]{Tielens:1983}
{Tielens} A.~G.~G.~M.,  1983, \aap, 119, 177

\bibitem[\protect\citeauthoryear{{Tielens} \& {Hagen}}{{Tielens} \&
  {Hagen}}{1982}]{Tielens:1982}
{Tielens} A. G. G.~M.,  {Hagen} W.,  1982, \aap, 114, 245

\bibitem[\protect\citeauthoryear{{van de Hulst}}{{van de
  Hulst}}{1949}]{Hulst:1949}
{van de Hulst} H.~C.,  1949, {The solid particles in interstellar space}.
Drukkerij Schotanus \& Jens, Utrecht

\bibitem[\protect\citeauthoryear{van Dishoeck, Herbst \& Neufeld}{van Dishoeck
  et~al.}{2013}]{Dishoeck:2013}
van Dishoeck E.~F.,  Herbst E.,    Neufeld D.~A.,  2013, Chemical Reviews, 113,
  9043

\bibitem[\protect\citeauthoryear{Vasyunin \& Herbst}{Vasyunin \&
  Herbst}{2013}]{Vasyunin:2013}
Vasyunin A.~I.,  Herbst E.,  2013, Astrophys.~J.~, 762, 86

\end{thebibliography}

\end{document}